\documentclass[printer]{mytTRB2e}
\usepackage[utf8]{inputenc}
\usepackage[T1]{fontenc}
\usepackage[english]{babel}
\usepackage{lmodern}
\usepackage[colorlinks,linkcolor=black,citecolor=black]{hyperref} 
\usepackage{booktabs}
\usepackage{units}
\usepackage{array}
\usepackage{xspace}
\newcommand{\ie}{i.\,e.\xspace}
\newcommand{\eg}{e.\,g.\xspace}

\newcommand{\expVal}[1]{\operatorname{E}(#1)}
\newcommand{\veps}{\varepsilon}
\newcommand{\lag}{\mathcal{L}_n}
\newcommand{\cs}[1]{\mathcal{R}_n^{#1}}
\newcommand{\lcur}{\ell_{\mathrm{cur}}}

\DeclareMathOperator{\cov}{Cov}
\DeclareMathOperator{\var}{Var}

\begin{document}

\title{A Microscopic Decision Model for Route Choice\\and Event-Driven Revisions}

\author{M.~Rausch$^{\ast}$\thanks{$^\ast$Corresponding author. Email: rausch@vwi.tu-dresden.de \vspace{6pt}}, Martin Treiber and Stefan Lämmer \\\vspace{6pt}
\emph{Technische Universität Dresden, Würzburger Str. 35, 01187 Dresden, Germany}
}

\maketitle


\begin{abstract}
We propose a microscopic decision model for route choice based on discrete choice theory. The correlation of overlapping routes is included in the random portions of the utility explicitly. For computational efficiency, we restrict the choice set to the turning possibilities at the next intersection, assuming shortest paths to the destination afterwards. The proposed decision model also regards traffic conditions (\eg~traffic lights, long queues) such that drivers may revise their previously taken decision. Due to its compatibility to already existing microscopic traffic flow models, the proposed route choice model can be readily simulated with available software. Combined, the proposed decision model features realistic behavior, \ie adaptive choice based on incomplete information and simultaneously allows for a straightforward implementation.
\end{abstract}

\section{Introduction}
The route choice of drivers has drawn much attention to researchers and practitioners in the past years. At the core of traffic assignment procedures, route choice models represent the behavioral assumptions that driver's decisions underlie. Most of the proposed route choice models are embedded in traffic assignment procedures searching for (stochastic) user equilibrium states.

In this paper, we develop a new microscopic route choice model that reflects the natural shortsightedness of drivers by assuming that only local knowledge, inferred by observations, and global assumptions for the traffic conditions in the network are included in decision processes. As decision processes are repetitive, the proposed model also allows drivers to dynamically revise their previously taken decision if spontaneous traffic events, where jams form and travel time increases, have occurred. To this end, the choice set is constrained to contain just the possible turning alternatives at the next node assuming shortest paths afterwards. In contrast to classic assignment approaches, however, no dynamical user equilibriums are calculated as the underlying behavioral assumptions are not realistic for spontaneous traffic disruption. Additionally, the proposed model allows for correlation among alternative routes and, thereby, addresses the daunting problem of overlapping routes. With its flexible and efficient formulation, the proposed route choice model is primarily intended to describe driver's route choice behavior under spontaneous traffic events. It complements the behavioral aspects of microscopic traffic models, including car-following, lane-change and signal-response models by a realistic route choice component.

The proposed model is stochastic and adheres to discrete choice theory in which all known characteristics of alternatives constituting a specified choice set are described by a deterministic utility~$ V $. However, since researchers cannot observe all factors influencing an individual's decision, a random utility portion~$ \varepsilon $ is introduced in order to capture these unobservable factors. In any case, the decision maker chooses the alternative with the maximum total utility  $ U = V + \veps $. However, since $ U $ is a random variable to the researcher, discrete choice models predict choice probabilities $ P $ for any alternative rather than fixed decisions. The statistical specification of the random utility $ \varepsilon $ determines whether the probability expressions $ P $ can be obtained in closed form or not. By choosing Gumbel-distributed random terms, one arrives at Logit choice models \citep{McFadden1974} that all have a closed form for $ P $. Contrarily, the Probit model \citep{Daganzo1979}, which allows correlation among alternatives, has Gaussian random error terms but is analytically not tractable. Although the Probit model is the most flexible random utility model, the Logit models are often given the preference due to their closed form.   

However, as the proposed route choice model follows an microscopic approach where driver's decision process is considered individually, no probabilities are calculated and the problem of a non-closed form of the Probit model is no longer relevant. Instead, each driver individually evaluates and maximizes their utility by choosing the best route.

The paper is organized as follows: In Sec.~\ref{sec:lit-review}, we give a brief overview of the major challenges in route choice modeling. Subsequently, we develop the microscopic decision model in Sec.~\ref{sec:dec-model} and show how it addresses the described challenges. In an analytical and numerical study, we check plausibility considerations and contrast the proposed model to selected random utility models regarding overlapping paths. Moreover, we demonstrate the applicability of the proposed model in a real-size networking assessing an incident management strategy. Finally, in Sec.~\ref{sec:discussion}, we summarize and conclude our findings.

\section{Major Challenges in Route Choice Modeling}
\label{sec:lit-review}
Compared to most other applications of choice modeling, route choice poses a particularly more complex problem for the following reasons: Firstly, in dense areas such as urban road networks, one has to cope with the combinatorial multitude of possible paths. Hence, techniques for generating path choice sets require appropriate approaches that extract relevant and reasonable subsets of all possible paths. Secondly, the problem of overlapping paths requires a proper treatment of the ensuing correlations in the utility functions. Finally, as vehicular traffic is dynamic by nature, route decisions may be dynamically revised if drivers face spontaneous traffic events.

\subsection{Path Set Generation}
The combinatorial multitude of available paths between an origin~$ O $ and a destination~$ D $, especially in urban networks, poses a major challenge for path set generation~\citep[\eg][]{Bovy2009}. Even computationally it seems impossible to consider choice sets containing all possible paths for every OD pair. In order to circumvent this \emph{path enumeration problem}, a variety of path set generation techniques were developed in the past. Principally, one can distinguish between implicit and explicit path enumeration techniques.

The approach to explicitly enumerate paths typically relies on the idea to calculate a number of~$ k $ shortest paths of an OD pair. Thereby, classical shortest paths algorithms are applied in an iterative manner. A common representative is the link elimination approach~\citep{Azevedo1993} that produces a set of ranked shortest paths by partly eliminating links from the shortest path found in the previous iteration. In the link labeling approach~\citep{Ben-Akiva1984} shortest paths regarding various objectives (e.\,g. travel time) are iteratively generated. Furthermore, \citet{DelaBarra1993} introduced the link penalty method where, instead of eliminating links, the link weights of the found shortest paths are systematically increased. By randomly increasing link weights after every determination of a shortest path, a conceptually similar approach was introduced by~\citet{Sheffi1982}. However, as pointed out by~\citet{Prato2009}, $ k $ shortest paths techniques are limited, \eg by the possibility of producing paths that are too similar. 

Recently,~\citet{Fosgerau2013} proposed a link-based Logit type model with an unrestricted choice set. Paths are generated implicitly as drivers choose next links on every node. Originally, this concept was introduced by~\citet{Dial1971} who required next links to (i) have a non-zero probability for being taken and (ii) do not backtrack, \ie that every next node of the path will bring the driver closer to the destination. Consequently, paths constitute implicitly as result of a transition from node to node during traffic assignment.

Another recent approach is the importance sampling~\citep{Frejinger2009, Flotterod2013} for which it is assumed that choice sets contain all possible paths in order to prevent biases due to a misspecification. Put simple, although only a subset of all possible paths is generated, a sample correction is evaluated and applied.  

\subsection{Overlapping Routes}
The consideration of overlapping routes in the choice process has challenged many researchers in the past. In order to capture the correlations among overlapping routes, researchers proposed to either include a correction term in the deterministic utilities or to introduce correlation in the random error terms explicitly by properly specifying their probability distributions. While both approaches have been utilized, model estimates suggest that the latter approach seems preferable~\citep{Bekhor2009}. 

While the classic multinomial Logit model (MNL) is not able to capture correlations among alternatives, the multinomial Probit model (MNP) applies for one of the most flexible discrete choice models. However, it lacks a closed analytical form for the choice probabilities making its use significantly more complicated. Retaining the simple form of a MNL model, \citet{Cascetta1996} and \citet{Ben-Akiva1999} developed the Commonality-Logit (CL) and Path-Size Logit (PSL) models, respectively. Both models assign each route a factor proportional to the overlap of the considered route to all other routes of the same OD relation. The average effect of the correlations is then attributed to the deterministic parts reducing them proportionally to the overlap. In effect, this allows to treat all routes obeying the same OD relation as independent options of a classic MNL model. Nevertheless, utilizing an MNP model, \citet{Yai1997} introduced a structured covariance matrix in which the overlap of alternative routes is treated by defining a random utility portion whose variance depends on common lengths of two overlapping routes. In this way, random utilities of overlapping routes are automatically correlated. 

An opponent approach to properly treating correlations among overlapping routes can be found in the group of structurally more complex models, that is, the General Extreme Value (GEV) model class proposed by~\citet{McFadden1978}. While the choice set structure becomes more complex, random utility terms are often chosen to be Gumbel deviates. In these cases, the models adhere to a Logit type model and, consequently, have a closed form. The most common representative of the GEV model class is the Nested-Logit-Model (NL)~\citet{Ben-Akiva1973, Ben-Akiva1974}, which was adapted to route choice, for instance, by~\citet{Abdel-Aty1998}. However, \citet{Prashker2004} states that NL models are not suitable for route choice modeling due to their constraint for alternatives to appear only in one nest.

Circumventing this constraint of the NL model,~\citet{Vovsha1997} developed the Cross-Nested-Logit model (CNL) in which alternatives are allowed to appear in more than one nest. Specifically adapting the CNL towards route choice,~\citet{Vovsha1998} proposed the Link-Nested-Logit model (LNL) in which every link in the network represents a nest holding all routes traversing through that link. Another representative of the GEV model class is the Paired-Combinatorial-Logit (PCL) model~\citep{Chu1989, Koppelman2000} that relaxes the constraints even further by allowing each pair of alternatives to have its individual covariance of random utilities. Ultimately, \citet{Wen2001} developed the Generalized-Nested-Logit (GNL) which includes the LNL and PCL models as special cases.

More recent models addressing route overlapping include the Link-Based Joint Network GEV model (LBJNG)~\citep{Papola2013} which describes a joint choice of all links constituting a route. Due to its link-based nature, route overlaps are explicitly captured. Furthermore, as it adheres to the GEV model class, the LBJNG model has a closed form for the choice probabilities. Another recent approach is the Path Size Weibit (PSW) model which was suggested by~\citet{Kitthamkesorn2013}. It assumes Weibull-distributed random terms and, analogously to the PSL model, captures correlation of overlapping routes in the deterministic via a path size factor. As introduced above, \citet{Fosgerau2013} developed a Recursive Logit (RL) model in which every node presents a point for the decision of the respective outgoing links. Route overlaps are captured by introducing a link size factor which, analogously to the PSL and PSW model, is attributed to the deterministic utilities of overlapping routes.

\subsection{Spontaneous Traffic Events and Disruptions}
Modeling the route choice behavior under spontaneous traffic events, which often pose disruptions, esp. in urban networks, is vital for the evaluation of counteracting management strategies. However only few attention has been attributed to the transient traffic flow patterns in spontaneously and short-termed disturbed networks. It is questionable if a new user equilibrium establishes under the disturbance~\citep[see also][]{Li2008}.

While travel experience from previous days is not beneficial for spontaneous traffic events, local observations, that is temporally and spatially constrained \emph{new} travel experience, is the only non-trivial information that drivers can infer from their environments. Recently, microscopic and mesoscopic traffic simulation environments have been developed in order to evaluate and analyze traffic flows in networks. However, most of the simulation models try to seek equilibrium states which is unrealistic for disturbed traffic flow patterns.  

\citet{Ben-Akiva2010} introduced \textsc{MitSim}Lab where a PSL model is applied in order to model route choice behavior in a link-based fashion. Within \textsc{MitSim}Lab, informed and uninformed drivers are considered, the first having a more certain perception of travel times in the network. PTV Vissim~\citep{Fellendorf2010} applies an iterative dynamic assignment technique measuring travel times on all links over predefined time intervals which together give the simulation duration. Once one interval has passed, the evaluated travel times are included in the decision processes in the next iteration. \textsc{Mezzo}~\citep{Burghout2004} is a mesoscopic simulation tool which allows informed drivers to switch better routes in case of spontaneous traffic events. A very recent approach was proposed by~\citet{Sanchez-medina2015} which allows drivers to adapt route choice at every intersection in the network and so accounts for dynamical changes of traffic conditions in the network.

\section{The Event-oriented Decision Model}
\label{sec:dec-model}
We propose a microscopic route choice model in which individual drivers make decisions about the possible turning directions at the next node, thereby observing traffic conditions in their local environments. Afterwards, a shortest path, on which traffic conditions remain uncertain, is associated to every possible turning direction leading to driver's destination. Individual decision processes are traversed in a repetitive manner such that, during their journeys, modeled drivers are able to revise their previously taken choice based on local observations. The proposed model is based on discrete choice theory and adapts the structure of a standard multinomial Probit model with normally distributed random utility terms.

\subsection{Choice Set}
The choice set~$ \cs{k} $ of driver~$ n $ includes shortest paths for all turning possibilities at node~$ k $ including going straight ahead and turning around, if allowed. Hence, every turning possibility at node~$ k $ is associated with the remaining route $ r $ to driver's destination. Hence, the choice set~$ \cs{k} $ is generally restricted to four alternatives on standard cross-intersections. This specification of the choice set is a distinctive feature of the proposed model since it drastically reduces the complexity: Generally, the choice set would include the full combinatorial multitude of all possible routes from a given origin or node to the destination.

\begin{figure}[tb]
\centering
\includegraphics[width=0.9\textwidth]{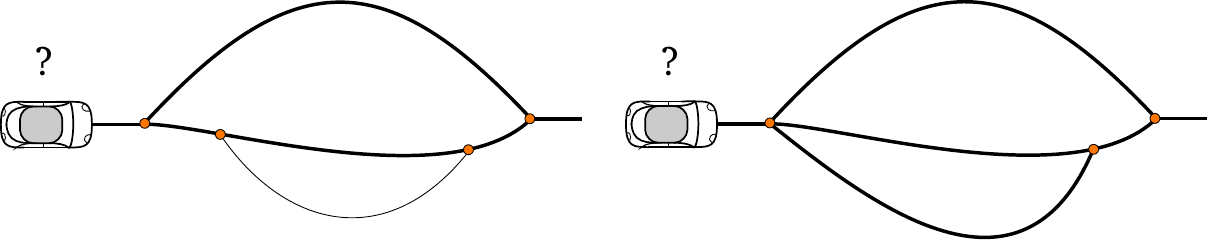}
\caption{\emph{Choice set}: The choice set of a driver at a decision point includes the shortest path (bold) of every respective turning direction at the next node. In the left situation, the choice set of the driver contains exactly two paths, whereas in the right frame, the next node has three turnings such that the driver can choose among 3 possible routes.}
\label{fig:routeSketch} 
\end{figure}

Shortest paths from driver's current node $ k $ to their destination are calculated with respect to a weighted graph formulation of the network. Typical link weights comprise link lengths or expected travel times under undisturbed traffic conditions. The major accelerating factor of this model is that the path calculation is done \emph{prior} to the simulation run for every node-destination combination. By taking the undisturbed network as basis, the shortest paths remain always the same. Hence, no additional paths have to be determined during the simulation run. Nevertheless, the choice set of any driver is dynamic in that it includes new alternative routes whenever driver~$ n $ passes node~$ k $ and enters a new road segment with end node~$ j $. The updated choice set~$ \cs{j} $ now includes all turning possibilities of node~$ j $ only retaining driver's currently chosen route. See Fig.~\ref{fig:routeSketch} for a depiction of the choice set.

\subsection{Decision Processes}
In discrete choice theory, a decision maker~$ n $ evaluates the utility $ U_{nr} $ of each alternative $ r $ and subsequently chooses the one with highest utility, that is alternative $ q $ for which holds
\begin{align}
U_{nq} > U_{nr} \quad \text{for all}\; r \in \cs{k}\; \text{with}\; q \not= r.
\end{align}
As any route~$ r $ of driver~$ n $ can be composed of a sequence of links $ \ell $, the corresponding utility~$ U_{nr} $ is obtained by
\begin{equation}
U_{nr} = \sum_{\ell \in \lag(r)} \left( V_{n\ell} + \varepsilon_{n\ell} \right),
\label{eq:general-route-utility}
\end{equation}
where $ \lag(r) $ denotes the set of all remaining links of driver's route~$ r $, $ V_{n\ell} $ the deterministic and~$ \varepsilon_{n\ell} $ the random utility of link~$ \ell $ perceived by driver~$ n $.

\subsubsection{Route Revisions}
During simulation, drivers may reassess their alternatives in the sense of Eq.~(\ref{eq:general-route-utility}) while permanently observing local traffic conditions. Combined, this does not only represent the natural short-sightedness of human drivers but also enables switches to alternative routes if obstructions are noticed along the currently chosen route. Revisions have to be triggered during the journey of modeled drivers. To this end, we propose three different trigger scenarios in the following which can also be combined. 

Notice that the driver has a complete route to the destination at any time, so it is not required to take a new decision whenever the set of alternatives changes due to the passage of the next node.

\begin{itemize}
\item \emph{Time-based Revisions}\\
In this scenario, it is assumed that there exists a random time interval $ T $ from one to the next decision process. Without loss of generality, we suppose that time intervals are independent with respect to different drivers and with respect to past time intervals of a given driver. For instance,~$ T $ can be chosen to be independently and identically exponentially distributed such that 
\begin{equation}
\mathrm{Prob}(T\le t)=1-\mathrm{e}^{-t/\tau}.
\end{equation}
Here,~$ \tau $ plays the role of a model parameter and reflects the average time interval between two decision. Plausible values for~$ \tau $ lie around \unit[10-30]{s}. 

\item \emph{Location-based Revisions}\\
In the location-based scenario, the onset of individual decision processes are tied to certain locations in the network. For instance, a revision process can be triggered whenever drivers enter new road segments. In analogy to time-based revisions, it is possible to trigger decision processes at random locations in the network.

\item \emph{Event-based Revisions}\\
Ultimately, decision processes can be triggered in conjunction with traffic events, for instance, when a driver encounters obstructions along their route (\eg incidents or traffic jams). However, in order to be consistent with reality, the triggering traffic event has to be \emph{unexpected}, that is, unplanned from the driver's perspective. This would be the case if, for instance, the driver unexpectedly encounters a standstill for longer than 3 minutes or receives some information about a traffic event in the network.
\end{itemize}

Note that a decision process must not necessarily result into the choice of a new route for the driver. This is only the case if the utility of any route is higher than the originally chosen one. In consistence with discrete choice theory, one can argue that (re)choosing the current route is also a decision. 

\subsection{Deterministic Utilities}
In contrast to many other models, the deterministic utility within the proposed event-orientied route choice model is measured in units of time. Particularly, the deterministic utility is the negative remaining travel time~$ T_{nr} $ that driver~$ n $ expects for route~$ r $.\footnote{More in line with the standard formulation, one could also define the deterministic link utility by $ V_{\ell} = \beta T_{\ell} $ where the model parameter $ \beta < 0 $ denotes the time sensitivity. Then, the variance of the random link utility is parameter free and can be defined by $ \sigma^2_{\ell} = L_{\ell} / L_0 $ where $ L_0 $ is the unit length (\eg $L_0=\unit[1]{km}$). While these two formulations are equivalent (if $ \lambda = 1 / (L_0 \beta^2) $), we consider our formulation as more succinct for the microscopic approach.} For the travel time estimation, modeled drivers consider their current links~$ \lcur $ and all remaining links~$ \ell \in \lag(r) $ to the destination of route~$ r $ separately. Whereas observed local traffic conditions go into the travel time estimation for the current link, assumptions are made by the drivers for all remaining non-observable links to the destination.

\subsubsection{Local Observations}
The only non-trivial traffic information available to the driver is inferred by the observation of local traffic conditions from the driver's location on the current link up to the end node, which should generally be perceptible in urban networks. Observable quantities on the current link include, for instance, the queue length $ d_{nr}^{\rm que} $, the current signaling $ S_{nr} $ and, based on the driver's link coordinate $ x_{nr}^{\lcur} $, the remaining distance to the end node of the current link. In general, this makes up for a estimation function $ \hat{T}_{\lcur}(\cdot) $ that combines all considered observables and is given by
\begin{align}
T^{\lcur}_{nr} = \hat{T}^{\lcur}\left(x^{\lcur}_{nr}, d^{\rm que}_{nr}, S_{nr}, \dots\right),
\label{eq:event-charge-current-link}
\end{align}
This function poses an observation model and can be flexibly specified by the modeler. All quantities are observed locally and actually correspond to each turning directions~$ r $ at the intersection, respectively. For this formulation, note that each local turning possibility at a node corresponds to the shortest route~$ r $ heading towards the destination. In case of non-exclusive turning lanes, however, the travel time estimation for the current link applies to more than one turning direction and, accordingly, to more than one route accordingly.

\subsubsection{Assumptions for the Remaining Travel Time}
Typically, drivers are only able to observe a certain length fraction of the next links~$ \ell \in \lag(r) $ of route~$ r $, if at all, such that the traffic information for those links cannot be inferred completely. Hence, we assume that for the remaining links~$ \ell \in \lag(r) $ of route~$ r $  driver~$ n $ makes assumptions about typical traffic conditions based on the undisturbed network. This results in a travel time estimation $ T_{nr}^{\mathrm{rem}} $ which is obtained by summing up the assumed travel times for all remaining links:
\begin{align}
T^{nr}_{\mathrm{rem}} = \sum_{\ell \in \lag(r)} \hat{T}^\ell(L_\ell, Q_\ell, \dots).
\end{align}
Likewise, assumptions are made by a generic estimation function $ \hat{T}^\ell(\cdot) $, considering, for instance, the length $ L_\ell $ or typical traffic flows $ Q_\ell $ of link $ \ell $.  

\subsubsection{Driver Persistence}
In order to prevent fast alternations of route choice decisions, the deterministic utility of the currently chosen route~$ q $ includes a positive bias $ V_0 $ in form of an alternative-specific constant representing a route switching threshold. This reflects the steadiness of a taken decision in the following sense: A new route is only chosen if the utility for it is larger by at least $ V_0 $ (\unit[1-5]{min}) compared to the actual route.

\begin{figure}[tb]
\centering
\includegraphics[width=0.8\textwidth]{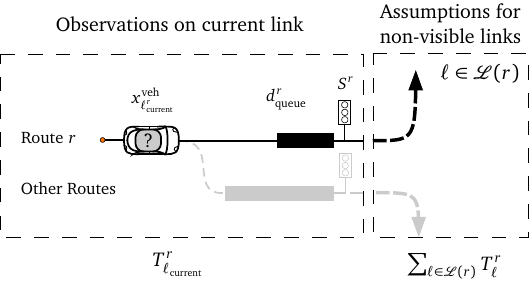}
\caption{\emph{Deterministic Utility}: The deterministic utility perceived by driver~$ n $ is the negative expected travel time $ T_{nr} $ for a route~$ r $ which is inferred by (i) observations on the current link and (ii) assumptions for links the driver has not yet entered. In the shown situation, the driver~$ n $ follows route~$ r $ while observing the local traffic conditions (such as signaling $ S_{nr} $ and queue length $ d_{nr}^{\rm que} $) for all turning possibilities, including the chosen one. Assumptions are then made for any non-observable link of remaing to the destination.}
\label{fig:detUtility} 
\end{figure}

Combined, in the proposed event-oriented route choice model, the deterministic utility~$ V_{nr} $ of route~$ r $ perceived by driver~$ n $ comprises the expected travel time required (i) on the current link $ \ell_{\rm cur}^r $ and (ii) for the not yet entered remaining links $ \ell \in lar(r) $ of route (turning direction) $ r $. Ultimately, this amounts to
\begin{align}
V_{nr} = -T_{nr} = -T_{nr}^{\lcur} - T_{nr}^{\mathrm{rem}} + \delta_{rq}\,V_0,
\end{align}
where $ \delta_{rq} $ is 1 if $ r = q $ and 0 otherwise. See also Fig.~\ref{fig:detUtility} for an illustration of the specification of the deterministic utilities in the proposed event-oriented route choice model.

\subsection{Random Utilities}
Random utility terms in the proposed event-oriented route choice model are distributed normally. However, rather than specifying a covariance matrix for the Gaussian distributions, we include the correlation structure in the random utilities explicitly. This is done by assuming link-based random utility terms~$ \veps_{n\ell} $ that are unique to every individual driver~$ n $ over the whole simulation runtime. The variance of the random utility~$ \veps_{n\ell} $ is proportional to the link length~$ L_{\ell} $:
\begin{equation}
\veps_{\ell} \sim \mathcal{N}\left(0, \sigma^2_{\ell}\right), \quad
\sigma^2_{\ell} = \lambda L_{\ell}.
\label{eq:epsilon-link}
\end{equation}
Here,~$ \lambda $ is a variance density representing the density of exogenous factors contributing to the stochastic utility, \eg bakeries, gas-stations or other points of interests, which might be attractive to the driver or not and personal preferences. 

All link random utilities are perfectly uncorrelated with respect to different links and different drivers and they are constant in time. The latter reflects the fact that, for instance, bakeries and gas stations are stationary and that personal preferences do not change on the time scale of the journey. Consequently, for each driver~$ n $, the random utilities~$ \veps_{n\ell} $ of \emph{all} links of the network are drawn from independent Gaussian distributions, whenever this driver enters the network. This drawing is accomplished by an efficient mapping
\begin{displaymath}
\left({\rm id}_{\rm Driver},\,{\rm id}_{\rm Link}\right) \rightarrow z \sim \mathcal{N}\left(0, \sigma^2_{\ell}\right)
\end{displaymath}
that quickly generates driver-link-unique random numbers and, thus, results in a fast simulation. 

Notice that the random utilities contain factors that are observable by the driver and are measured in units of time. This requires to choose the variance density~$ \lambda $ in units of, e.\,g., $ [\unit{min}^2/\unit{km}] $. A plausible parameter value can be given by assuming that, over a distance of \unit[10]{km}, the driver preferences unknown to the researcher accumulate to a standard deviation of \unit[10]{min} leading to $ \lambda = \unit[10]{min^2/km} $ (cf. Table~\ref{tab:model-parameters}). Find all proposed model parameters summarized in Tab.~\ref{tab:model-parameters}.

\begin{table}[tb]
\centering
\caption{Taken together, our proposed decision model features three basic model parameters. Depending on the formulation of the link weights for the undisturbed situation (e.g. by capacity-restraint functions), and on the heuristic rules used to assess the impacts of incidents, further parameters could be necessary to completely specify the model.}
\begin{tabular}{cm{0.53\textwidth}m{0.17\textwidth}}
\toprule
Parameter & Description & Typical value\\
\midrule
$ \lambda $ & Variance density representing the density of
individual preferences and points of interest & \unit[5-10]{\nicefrac{min$^2$}{km}}\\
\midrule
$ V_0 $ & Persistence component of the deterministic utility valid for the currently chosen route of driver $ n $. It characterizes the steadiness of a taken a decision. & \unit[1-5]{min}\\
\midrule
$ \tau $ & Average time interval between two decisions & \unit[10-20]{s}\\
\bottomrule
\end{tabular}
\label{tab:model-parameters}
\end{table}

\section{Model Properties and Plausibility Considerations}
In the following, we will show that the above specification of the random utilities satisfies two consistency conditions that should be required for any plausible specification. 

\subsection{Split-Link Consistency}
If one splits a link $ \ell $ of length $ L $ into two parts $ \ell_{a} $ and $ \ell_{b} $ of lengths $ L_{a} $ and $L_{b}$ (with $ L_{a} + L_{b} = L $), respectively, by introducing a degenerate node, the statistical properties of the random utility of the partial route consisting of the links $\ell_{a} $ and $ \ell_{b} $ should be identical to that of the undivided link  $\ell $. Our specification satisfies the split-link consistency since the sum of two independent Gaussians of expectation zero and variances $ \lambda L_{a} $ and $ \lambda L_{b} $, respectively, gives a Gaussian of expectation zero and variance $ \lambda(L_{a} + L_{b}) = \lambda L $ which is the statistical specification of the random utility of the undivided link.

\subsection{Automatic Correlation of Overlapping Routes}
Consider two overlapping routes $ r $ and $ q $. Due to the link additivity of the random utilities, it is always possible to decompose the random utility of a route into (i) a part that only contains the random utilities from unshared links, $ \veps^{\rm u}$, and (ii) a part that only contains the random utility of shared links, $ \veps^{\rm s} $. Consequently, the random utilities of the routes $ r $ or $ q $, consisting each of some unshared and shared links, are given by
\begin{equation}
\label{eq:sharedLinks}
\veps_{r/q} = \veps_{r/q}^{\rm u} + \veps_{\mathit{rq}}^{\rm s}.
\end{equation}
Here, $ r/q $ refers to either $ r $ or $ q $, whereas $ \mathit{rq} $ refers to the common parts of both routes. Apparently, the same decomposition can be done for the length $ L_r $ of a route $ r $:
\begin{displaymath}
L_{r/q} = L_{r/q}^{\rm u} + L_{\mathit{rq}}^{\rm s}.
\end{displaymath}

In the previous section, we claimed that overlapping routes are correlated by construction, as the random utilities are driver-link-based. This can be seen by explicitly calculating the covariance of the random utilities of the two routes $ r $ and $ q $ specified in Eq.~\eqref{eq:sharedLinks}. The covariance amounts to
\begin{displaymath}
\cov(\veps_{r}, \veps_{q})
= \cov(\veps_r^{\rm u} + \veps_{\mathit{rq}}^{\rm s},
\veps_q^{\rm u} + \veps_{\mathit{rq}}^{\rm s}) 
= \expVal{(\veps_{\mathit{rq}}^{\rm s})^2}
= \var(\veps_{\mathit{rq}}^{\rm s}) = \lambda \cdot L_{\mathit{rq}}^{\rm s}
= \lambda \sum_{\ell \in \mathcal{L}^{\rm s}(\mathit{rq})} L_{\ell},
\end{displaymath}
where we have used the definition~\eqref{eq:epsilon-link} for the variance density $ \lambda $ and $ \mathcal{L}^{\rm s}(\mathit{rq}) $ is the set of shared links $ \ell $ of routes $ r $ and $ q $. This results in the correlation coefficient
\begin{align}
\rho_{\veps_{r}, \veps_{q}} = \rho_{\mathit{rq}} = \frac{\cov(\veps_{r}, \veps_{q})}{\sqrt{\var(\veps_{r})} \sqrt{\var(\veps_{q}})} = \frac{L_{\mathit{rq}}^{\rm s}}{\sqrt{L_r L_q}}.
\label{eq:correlation-coefficient}
\end{align}

\begin{figure}[tb]
\centering
\includegraphics[width=0.6\textwidth]{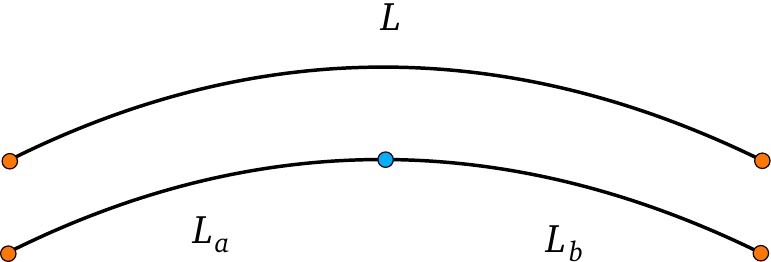}
\caption{\emph{Demonstration of the ``Split-Link Consistency''}: The statistical properties of a link with length $ L $ and its splitting into two parts of length $ L_a $ and $ L_b $, respectively, should be equivalent.}
\label{fig:splitLink}
\end{figure}

Generally, the correlation coefficient will never become negative as the overlap portion $ L_{\mathit{rq}}^s $ is always non-negative. Moreover, if, and only if, the routes $ r $ and $ q $ are identical, the correlation coefficient is equal to unity. Furthermore, Eq.~\eqref{eq:correlation-coefficient} reveals that the correlation coefficient is identical to the overlap fraction of routes $ r $ and $ q $ when using the geometric mean of the lengths of both routes. This becomes even more apparent if one considers the special case $ L_r = L_q = L $ yielding $ \rho_{\mathit{rq}} = L_{\mathit{rq}}^s / L $.

\section{Stochastic Loading in the Overlap Network---Numerical Findings}
In a well-known network~\citep{Sheffi1982}, we apply the proposed microscopic route choice model to a stochastic loading procedure. The obtained link flow pattern is contrasted to the loading patterns of the MNL, CL, PSL and LNL models.

\begin{figure}[tb]
\centering
\includegraphics[width=0.6	\textwidth]{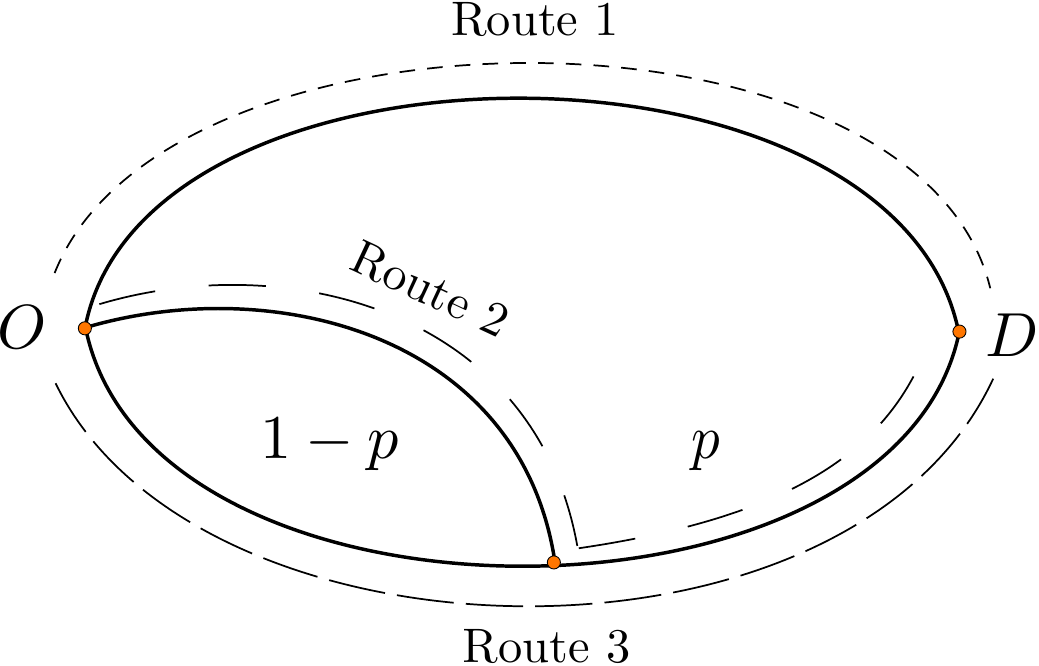}
\caption{In this well-known network~\citep{Sheffi1982}, one OD pair is connected by three distinct routes. Two routes, however, overlap by some portion $ p $. For simplicity, we assume that all routes have the same travel time.}
\label{fig:overlapping-network}
\end{figure}

The overlap network connects an OD pair by three routes. While Route 1 uses an exclusive link, Routes 2 and 3, are overlapping by some portion $ p $, see Fig.~\ref{fig:overlapping-network}. For the sake of simplicity, all routes are considered to have the same travel time $ T $. The deterministic utility is then chosen to be $ V = -T $. Hence, we do not consider any congestion effects in the overlap network, nor variable traffic conditions. This implies that the event part $ T_{\ell_{\rm cur}^r}^r\left(x_{\ell_{\rm cur}^r}^{\rm veh}, d_{\rm que}^r, S^r, ...\right) $ of the proposed model is set to zero.

We assume three distinct overlap manifestations: (i) $ p = 0.05 $ (minor overlap), (ii)  $ p = 0.5 $ (significant overlap) and (iii) $ p = 0.95 $ (nearly complete overlap). From the driver's perspective, Route 1 is always considered as distinct alternative. Depending on the overlap manifestation, this does not necessarily apply to Routes 2 and 3. In the first overlap manifestation, $ p = 0.05 $, the driver faces one fully and two nearly distinct routes. Under equal route properties, the route choice probability becomes approximately $ P =  \nicefrac{1}{3} $ for each of the three routes. Contrarily, in overlap manifestation three, $ p = 0.95 $, Routes 2 and 3 are nearly completely overlapped such that they represent, physically, only one route. Hence, the choice probabilities result in approximately $ P = \nicefrac{1}{2} $ for Route 1 and the set of Routes 2 and 3, respectively. Thereby, to each of the overlapping routes a probability near $ P = \nicefrac{1}{4} $ is allotted.

In the application of the proposed model, we have simulated 10000 drivers that individually went through decision processes at the end of which stood the realized route choice. For the variance density, we assumed $ \lambda = \unitfrac[5]{min^2}{km} $. In contrast, we have analytically calculated the choice probabilities for the MNL, CL, PSL, and LNL models. For the CL and PSL models, we employed the original formulations of the commonality factor and path size attribute, respectively. In the LNL model, we assumed a constant nesting coefficient $ \mu = 0.1 $. See Tab.~\ref{tab:overlap-results} for a result comparison.

As expected, the MNL does not take the overlap into consideration. The choice probabilities for the three routes remain exactly the same for all overlap manifestations, namely $ \nicefrac{1}{3} $ respectively. While revealing the limitation of the MNL, particularly in the case of nearly complete overlapping paths, the CL, PSL and LNL models predict more reasonable choice probabilities as stated above. The proposed microscopic route choice model employs a different approach in order to include the overlap-induced correlation but results in similar values with respect to the CL, PSL and LNL models. Note however that no probability expressions were calculated but the obtained values correspond to relative frequencies of chosen routes of the 10000 simulated drivers.

\begin{table}[tb]
\centering
\caption{Comparison of the relative frequency of the chosen routes in the proposed model with the choice probabilities of the MNL, CL, PSL and LNL models.}
\vspace{0.5em}
\begin{tabular}{ccccccc}
\toprule
& Route & Proposed Model & MNL & CL & PSL & LNL \\
\midrule
 			 & 1 & 0.342 & 0.333 & 0.344 & 0.338 & 0.338 \\
$ p = 0.05 $ & 2 & 0.333 & 0.333 & 0.328 & 0.331 & 0.331 \\
 			 & 3 & 0.325 & 0.333 & 0.328 & 0.331 & 0.331 \\
\midrule
 			 & 1 & 0.391 & 0.333 & 0.428 & 0.400 & 0.394 \\
$ p = 0.5 $  & 2 & 0.306 & 0.333 & 0.286 & 0.300 & 0.303 \\
 			 & 3 & 0.303 & 0.333 & 0.286 & 0.300 & 0.303 \\
\midrule
 			 & 1 & 0.475 & 0.333 & 0.497 & 0.488 & 0.472 \\
$ p = 0.95 $ & 2 & 0.271 & 0.333 & 0.253 & 0.256 & 0.264 \\
 			 & 3 & 0.254 & 0.333 & 0.253 & 0.256 & 0.264 \\
\bottomrule
\end{tabular}
\label{tab:overlap-results}
\end{table}

\section{Applications}
\begin{figure}[tb]
\centering
\includegraphics[width=0.65\textwidth]{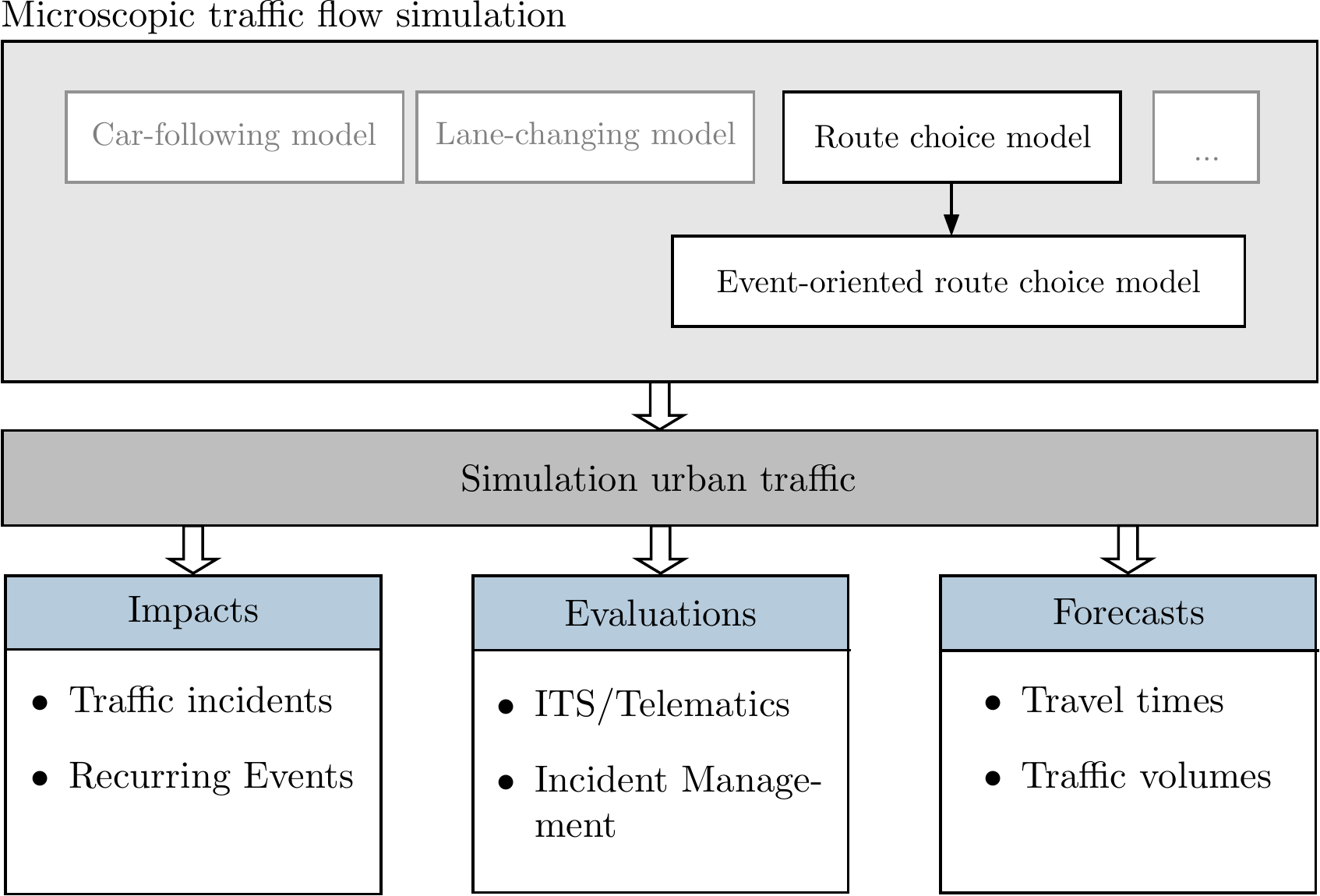}
\caption{Classification and application fields of the proposed event-oriented route choice model.}
\label{fig:applications} 
\end{figure}

The proposed route choice approach mainly focuses on modeling the short-termed driver behavior under presence of spontaneous, that is, unexpected traffic events where jams form and travel times increase. It realistically assumes that drivers anticipate unexpected congestion (\ie travel time increase) on basis of local observations in their visible environments or alternative, non-trivial, forms of information. In contrast to classic traffic assignment techniques, where drivers are able to instantly infer at least approximated actual travel times, in the proposed model driver assume only typical travel times for any non-observable portion of the network. Note that typical travel times in the network not necessarily correspond to actual travel times, esp. in case of traffic incidents. Based on the proposed model's features, its main application context is urban traffic. As it follows a microscopic approach, it complements the car-following, lane-changing and signal-responding models by a route choice component. As depicted in Fig.~\ref{fig:applications}, the model is especially applicable for estimating the impact of traffic events, for evaluating traffic management strategies and for creating online traffic forecasts for incident situations. 

\subsection{Evaluation of an Traffic Incident Management Strategy}
\begin{figure}[tb]
\centering
\includegraphics[width=0.65\textwidth]{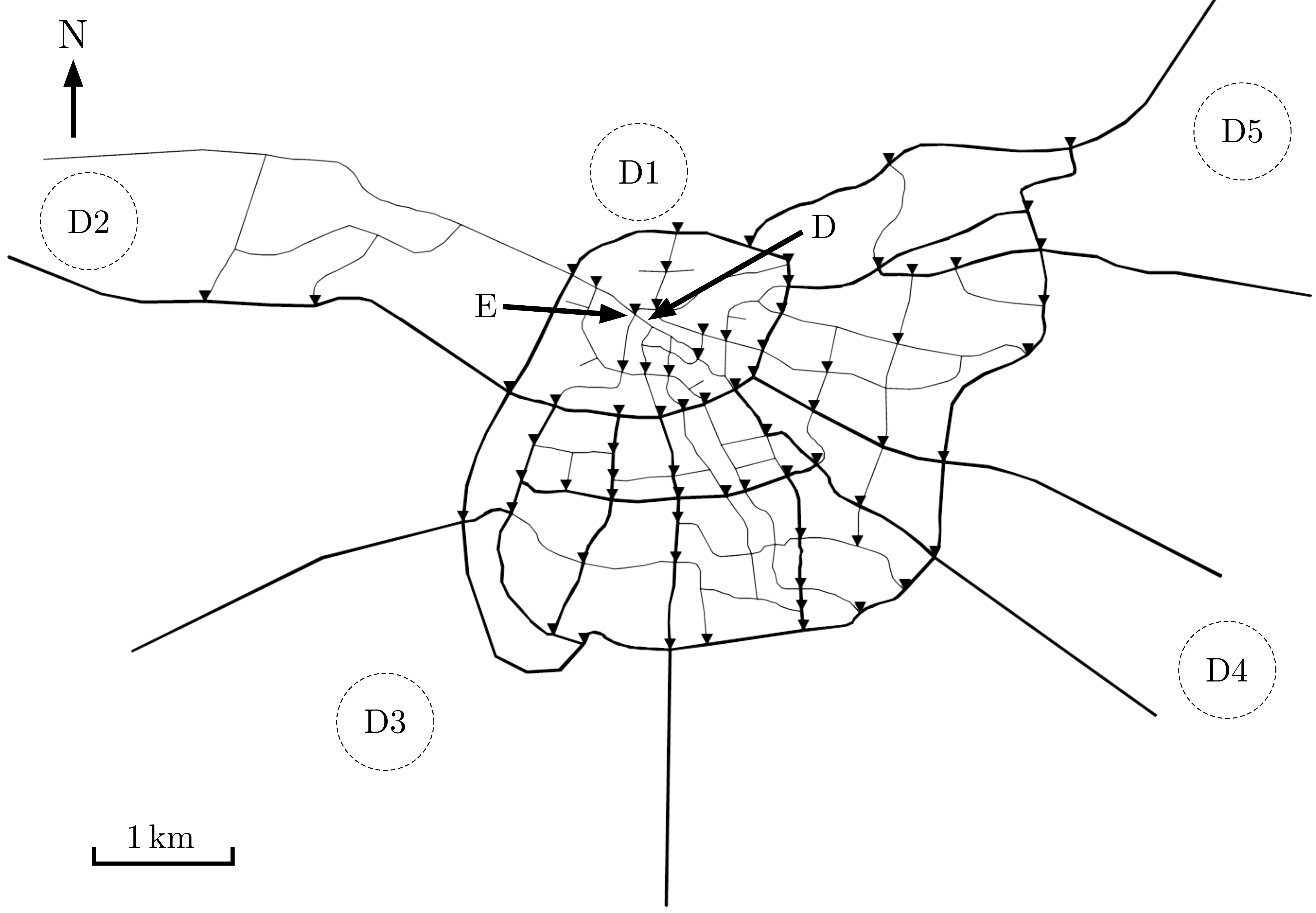}
\caption{Examined real-size network based on the French city Avignon. Triangles mark signal-controlled intersections, D1--D5 are traffic districts and E and D are locations where incidents were modeled.}
\label{fig:complexnetwork} 
\end{figure}

The following section gives an example for the application of the proposed decision model. Particularly, an innovative urban traffic incident management strategy, a self-organizing and local inflow regulation principle, proposed by the authors \citep{Lammer2013} is evaluated in the real-size Avignon network depicted in Fig.~\ref{fig:complexnetwork}. Put simple, it regulates the inflow into already congested road segments by restricting or skipping green times while, simultaneously, remaining green times are used for the yet free turning directions. In order to realistically incorporate this principle into route choice behavior of drivers, the the proposed model is employed under the assumption that drivers observe traffic light signals at intersection. Specifically, an observation model for red times is applied assuming that, with increasing red times for their current turning direction, drivers become more impatient by perceiving that alternative turning directions are simultaneously provided with green times, the driver might revise the currently chosen route in favor of an alternative one. In this way, a self-organized redistribution of traffic flows sets in, utilizing road capacities surrounding the congested area. Although the observation model does not rely on real data, it is consistently formulated in that not \emph{all} drivers \emph{instantly} change routes when affected by unusually long red times.

\begin{figure}[b]
\centering
\includegraphics[width=0.65\textwidth]{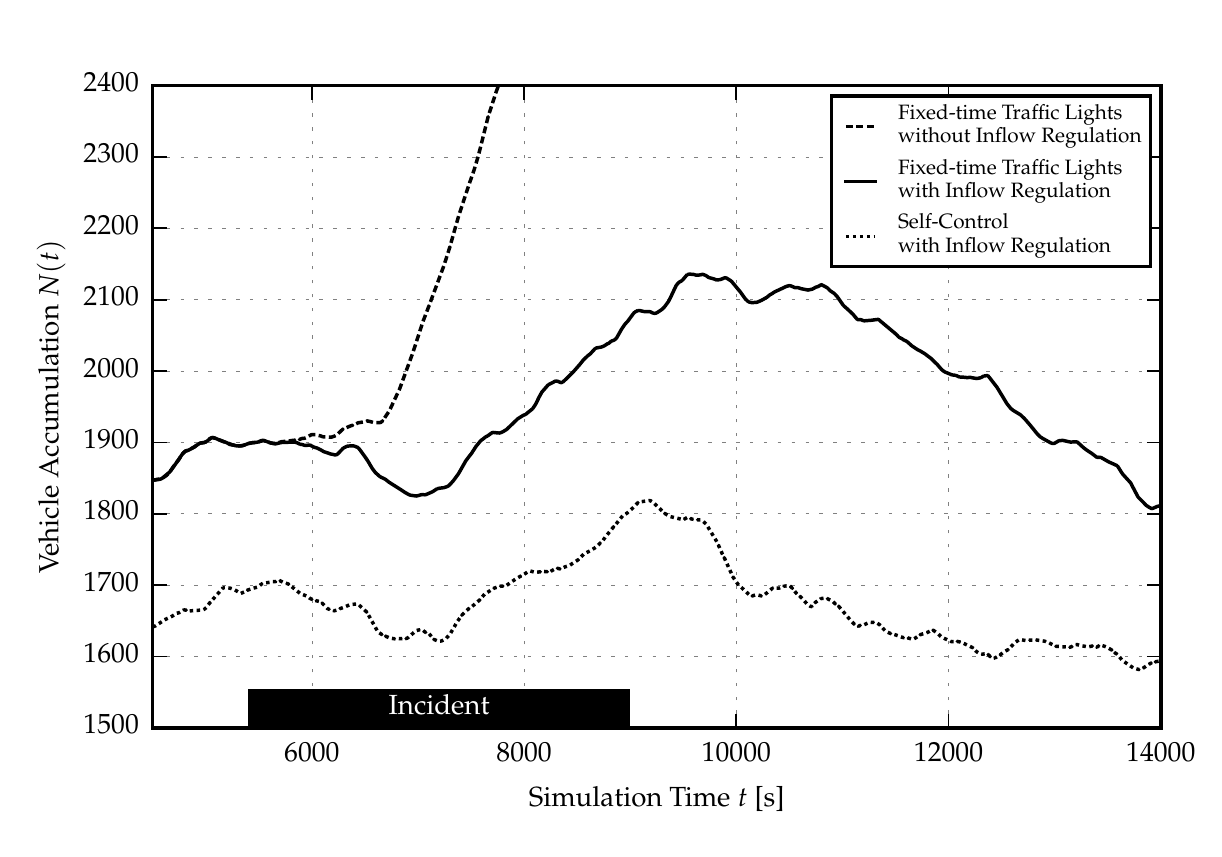}
\caption{Classification and application fields of the proposed event-oriented route choice model.}
\label{fig:complexnetwork_accumulation} 
\end{figure}

We modeled a typical urban incident resulting in stop and go waves at location D in the Avignon network (see Fig.~\ref{fig:complexnetwork}). For assessing the performance of the inflow regulation principle, we measured the vehicle accumulation in every time step. As simulation environment, we chose PTV Vissim (Version 5.4). The vehicle accumulation time series is shown in Fig.~\ref{fig:complexnetwork_accumulation}. We compared fixed time traffic lights with and without inflow regulation. Furthermore, the self-organizing Self-Control from \citet{Lammer2008} was considered with the inflow regulation principle. 

In all cases, the proposed route choice model was applied to every driver in the network. Drivers made their decisions on basis of local signaling observations and assumptions for non-visible network areas. Neither were they informed about actual travel times in the network nor did they know anything about the occurrence of the incident in advance. Apparently, applying the inflow regulation principle makes a distinct difference in terms of vehicle accumulation in the network: Much fewer vehicles are accumulated during the incident and gridlock effects are antagonized. Note that the efficiency of the proposed route choice model is a crucial feature: In the unregulated case (dashed line in Fig.~\ref{fig:complexnetwork_accumulation}) about 4\,000 drivers were simulated simultaneously on a typical personal computer with four CPU cores.

In a forthcoming paper, the authors will provide an thorough overview of described incident management strategy and perform a detailed evaluation study.

\section{Conclusions}
\label{sec:discussion}
In summary, the distinctive features of the proposed route choice model are (i) a simple choice set that reflects driver's shortsightedness with respect to dynamical changes, (ii) link-additive driver-based random utilities automatically leading to correlations between overlapping routes and (iii) repeated re-evaluation of dynamically changing choice sets allows the driver to react to variable local traffic conditions and, particularly, to incidents. Furthermore, the decision model is efficiently formulated such that its main context are microscopic traffic simulations. Put together, the proposed model is computationally simple and leads to a fast simulation while it is consistent with discrete choice theory. Due to its straightforward implementation, it easily complements the existing set of microscopic traffic models (car-following, lane-changing and signal-responding models) by a route choice component.

The main focus of the proposed model is to describe driver's route choice behavior under spontaneous traffic disruptions where the assumption of an existing user equilibrium is not realistic. Instead, drivers are assumed to base their decisions only on local observations and assumptions about travel times in the remaining network. The proposed model provides a broad spectrum of concrete applications especially in the field of Intelligent Transportation Systems (ITS). We have demonstrated its applicability for assessing a novel incident management strategy in a real-size network with thousands of drivers. In a future study, it could be interesting to see how the proposed model performs in an environment as used by \citet{Ben-Akiva2012} with DynaMIT-P.

Compared to other discrete route choice models, the inclusion of correlations between routes with shared links does not involve a route comparison (e.\,g CL, PSL models) or a more complex choice set structure (e.\,g. PCL, LNL models). Instead, the correlation is automatically implied by overlapping routes, which is (i) consistent with reality and (ii) accounts for a straightforward implementation of the decision model into route choice simulations. Our numerical findings in the overlap network revealed that the proposed model results in realized route choice fractions that are comparable to the CL, PSL and LNL models. Thus, the approach of using a variance density for random utilities is equivalent to already existing models.  

In the proposed model, we assume that the driver knows the shortest paths from any intersection to the destination, under normal conditions. While this poses a convenient model assumption, it might not hold for every driver in reality. On the other hand, shortest paths are precisely the information that navigation devices provide, so the model represents, in effect, drivers with a navigation system overriding their recommendations as a response to traffic incidents. As a result, drivers will not necessarily take the overall shortest route of their OD relation but dynamically adapt their previously taken route choices. Notice that some recent ``live'' navigation devices incorporate event-oriented traffic-dependent navigation. In principle, the (proprietary and non-disclosed) algorithms behind these systems tackle a similar problem as the proposed model.

There are several avenues to expand this research. As future prospectives, one could model informed drivers which might respond sooner to variable traffic conditions. Furthermore, in order to support long-term applications of the proposed model, one could implement a driver memory that will store empirical values for the link travel times instead of using the undisturbed network. 

\section*{Acknowledgement}
The authors thank the DFG (German Research Foundation) for partial
financial support of this research.

\bibliography{library}

\end{document}